\documentclass[twocolumn,twocolappendix]{aastex63}
\usepackage{nicefrac}
\usepackage{amsmath}
\usepackage{float}
\usepackage{bbold}
\usepackage{xspace}
\usepackage{placeins}
\usepackage[T1]{fontenc}
\usepackage{textcomp}
\usepackage[caption=false]{subfig}

\def\bhspin{a_*}
\def\ud{\mathrm{d}}
\def\qcarter{\mathcal{Q}}

\shortauthors{Wong}

\begin{document}

\title{Black Hole Glimmer Signatures of Mass, Spin, and Inclination}

\correspondingauthor{George~N.~Wong}
\email{gnwong2@illinois.edu}
\author[0000-0001-6952-2147]{George~N.~Wong}
\affil{Illinois Center for Advanced Studies of the Universe, Department of Physics, \\ University of Illinois, 1110 West Green St., Urbana, IL 61801, USA}
\affil{CCS-2, Los Alamos National Laboratory, P.O.~Box 1663, Los Alamos, NM 87545, USA}

\begin{abstract}  

Gravitational lensing near a black hole is strong enough that light rays can circle the event horizon multiple times.  Photons emitted in multiple directions at a single event, perhaps because of localized, impulsive heating of accreting plasma, take multiple paths to a distant observer.  In the Kerr geometry, each path is associated with a distinct light travel time and a distinct arrival location in the image plane, producing {\em black hole glimmer}.  This sequence of arrival times and locations uniquely encodes the mass and spin of the black hole and can be understood in terms of properties of bound photon orbits.  We provide a geometrically motivated treatment of Kerr glimmer and evaluate it numerically for simple hotspot models to show that glimmer can be measured in a finite-resolution observation. We discuss potential measurement methods and implications for tests of the Kerr hypothesis.

\end{abstract}


\section{Introduction} \label{sec:introduction}

Spinning supermassive black holes likely power astrophysical relativistic jets
via the Blandford--Znajek mechanism (\citealt{Blandford1977}, although see \citealt{Blandford1982} and \citealt{LyndenBell2006} for alternative jet-power mechanisms). To probe the spin--jet connection through observation, it is necessary to understand the properties of spacetime near the hole and to have an accurate model of the accreting plasma around the hole. If spacetime is described by the Kerr metric, then its properties are uniquely determined by the mass and angular momentum of the central black hole.

Many spin measurement methods propose measuring the size of the accretion disk, the qualities of disk oscillations, or the deviation in line profiles due to black hole spin \citep[e.g.,][]{hanawa1989xrayshiftspinmeasurement,Kojima1991spinlineprofile,laor1991lineprofile,kato2001qpospin,Miller2007reverberation}.
These measurement techniques rely on an accurate understanding of the accretion flow and are therefore subject to uncertainties in the plasma physics model. The 2017 Event Horizon Telescope observation of the black hole at the center of the galaxy M87 provided the first direct horizon-scale observation of a black hole \citep{EHT2019I}; however, its ability to constrain the spin of the hole is also limited by the modeling uncertainties. 

Strong gravitational lensing allows photons to travel along bound orbits that circle black holes (see \citealt{claudel2001photonsurface} for a treatment of bound photon surfaces in arbitrary spacetimes).
In the Kerr metric, the properties of these orbits are determined solely by the spacetime geometry, i.e., by the mass and angular momentum of the black hole, in the case of the vacuum metric. In the image plane, the asymptotically bound orbits produce a characteristic \emph{critical curve} whose size and shape are also set by the spacetime geometry.
Since the critical curve is determined solely by the spacetime, it is independent of the accretion model, and thus it provides a consistent signature that directly probes the hole's properties. 

Although the critical curve is not necessarily observable, it is often traced by a high intensity \emph{photon ring} that is produced as nearly bound orbits steadily sample the high-emissivity region near the hole (see \citealt{gralla2020curvering} for a discussion of some differences between signatures of the curve and ring, especially in the case of midplane emission).
Measurement strategies to infer bounds on spin from the shape of the ring or curve have been proposed in the past \citep[e.g.,][]{falcke2000shadowmeasure,Takahashi2004spinshape,Bambi2009,hioki2009spinshape,Younsi2016spinshadow,johnson2020universal}. Related proposals have suggested testing the hypothesis that spacetime is Kerr by looking for deviations in the shape of the curve
\citep[e.g.,][]{amarilla2010,Tsukamoto2014kerrandsimilar,Amarilla2013kaluzaklein,Mizuno2018Shadows,lia2020nonkerrshadow,Olivares2020,Wielgus2020}.

The bound orbits allow light near a black hole to orbit it multiple times before escaping to an observer. Since emitters can radiate in multiple directions simultaneously, two rays produced by the same source may orbit the hole a different number of times. Light signals from subsequent orbitings will be separated by a time delay set by the length of a complete winding around the hole. These delays will cause the source to echo in the image plane. Since the echo period is a function of path length, it is intrinsically tied to the underlying bound orbits and can be measured to infer spin.
Constraining spin by measuring dominant echoes has been considered in the past \citep[e.g.,][]{broderick2005imaginghotspots,moriyama2015spinmeasurement,Saida2017,Thompson2019Impulsive,moriyama2019spinvlbi,gralla2020criticalexp}.

In detail, the Kerr geometry produces a rich spectrum of echo time delays associated with resonant bound orbits. These resonant echo delays are closely related to the black hole quasinormal-mode spectrum in the eikonal limit \citep[see][]{Yang2012QNMKerrGeometry}, which has been studied in the context of gravitational-wave ringdown and measuring mass and spin \citep[e.g.,][]{berti2006qnmligo,Buonanno2007qnmequalmass,berti2007qnmunequalmass}.

Each echo maps to a distinct arrival location in the image plane; taken together, the set of echoes produces a characteristic \emph{black hole glimmer} 
that encodes the mass and spin of the hole. Since the mechanism that produces these echoes is driven purely by the spacetime, the black hole glimmer signature is separable from the source emission model. If glimmer can be measured precisely, it is possible to test the Kerr hypothesis and infer the black hole mass and spin, even without a detailed understanding of the emission source.
We provide a geometrically motivated treatment of Kerr glimmer and demonstrate that the glimmer signature of a hotspot can be measured even in a finite-resolution observation.

This article is structured as follows. In Section~\ref{sec:reviewkerr}, we review the salient features of the Kerr geometry in the context of bound orbits, and in Section~\ref{sec:photonshellobs} we describe the static observable image due to the bound orbits. We describe and explore the Kerr glimmer signature in Section~\ref{sec:echospectrum}, and we give a brief discussion and example measurement in Section~\ref{sec:discussion}. We summarize in Section~\ref{sec:summary}.

\section{Kerr geometry and bound orbits} \label{sec:reviewkerr}

Black holes (in vacuum) in general relativity are described by their mass, angular momentum, and charge, although it is unlikely that supermassive black holes with a dynamically important charge exist in nature. 
Charge-neutral holes with non-zero angular momentum are described by the Kerr metric. We use geometrized units with $G = c = 1$ and write angular momentum $J$ in terms of the conventional dimensionless spin parameter $\bhspin \equiv J/M^2$ where $M$ is the mass of the hole. Hereafter, we set $M=1$. In Boyer--Lindquist coordinates $x^\mu = \left( t, r, \theta, \phi\right)$, the Kerr line element is
\citep{Bardeen1972}
\begin{align}
    \ud s^2 =& - \left( 1 - \dfrac{2r}{\Sigma} \right) \, \ud t^2
    - \dfrac{4\, \bhspin r \sin^2 \theta}{\Sigma} \, \ud t \, \ud \phi + \dfrac{\Sigma}{\Delta} \, \ud r^2 \nonumber \\ 
    & + \Sigma \, \ud \theta^2 +  \dfrac{\left(r^2+\bhspin^2\right)^2 - \Delta \bhspin^2 \sin^2 \theta }{\Sigma} \sin^2 \theta \, \ud \phi^2 
\end{align}
with 
\begin{align}
    \Sigma \equiv r^2 + \bhspin^2 \cos^2 \theta, \qquad
    \Delta \equiv r^2 - 2r + \bhspin^2 \; .
\end{align} 

The Kerr geometry admits a set of spherical bound orbits---orbits with fixed radial coordinate---for both massive and massless particles near black holes. The set of all spherical massless/photon orbits is known as the \emph{photon shell} \citep{johnson2020universal}.
The observational consequences of the photon shell on black hole images have been studied both in the context of non-spinning holes \citep[see especially][]{luminet1979}
and the more general spinning Kerr hole \citep[e.g.,][]{bardeen1973kerrshadow}. One of the first full treatments of spherical photon orbits in Kerr was presented by \citet{Teo2003}, who also provided a convenient categorization of the types of such orbits. In this section, we review several key features of the bound photon orbits.
Appendix~\ref{sec:appendix:kerreom} details the results described in this section. We do not consider black holes with extremal spin.

For a given spin, the set of all bound spherical photons orbits can be parameterized by the radii of the orbits. These radii lie continuously in the range $r_- \le r \le r_+$, where
\begin{align}
    r_\pm = 2 \left( 1 + \cos \left( \dfrac{2}{3} \cos^{-1} \pm \bhspin \right) \right) .
    \label{eq:rminmax}
\end{align}
Only the two extremal orbits at $r_\pm$ are confined to the midplane. The prograde orbit at $r = r_-$ rotates around the hole in the same direction as its spin, and the retrograde orbit rotates opposite the spin at $r = r_+$. The other orbits at intermediate radii oscillate between symmetric minimum and maximum latitudes $\theta_{\pm}$.

\begin{figure*}[t!]
\begin{center}
\subfloat[Resonant orbits in 3d]{\begin{tabular}[b]{c}%
\includegraphics[width=0.2\textwidth]{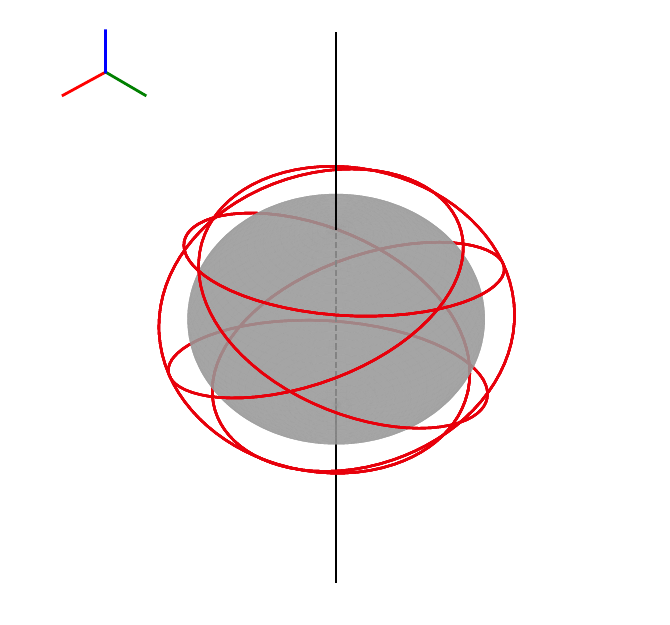} 
\includegraphics[width=0.2\textwidth]{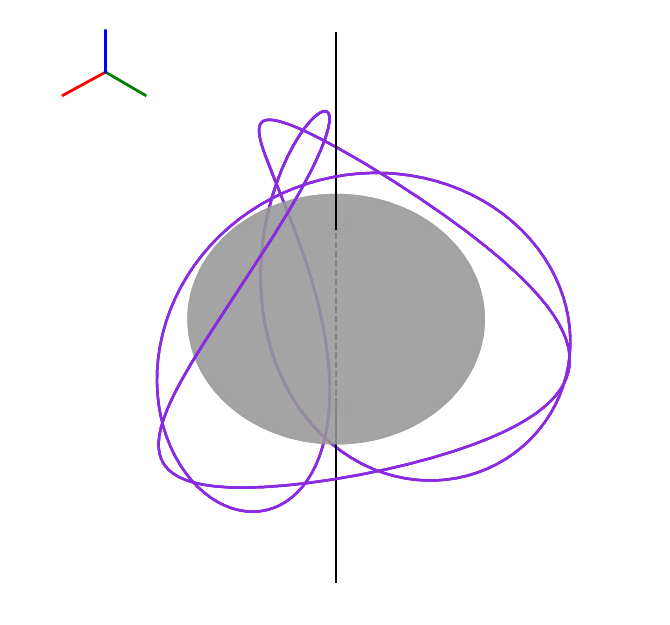} \\
\includegraphics[width=0.2\textwidth]{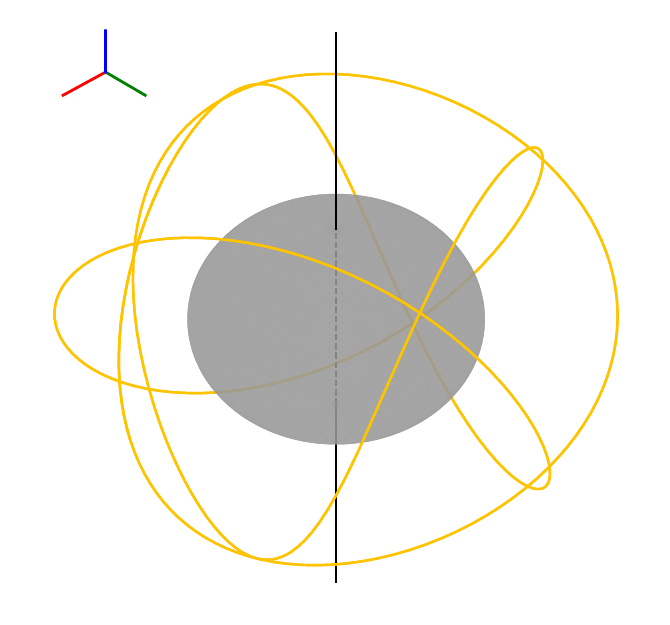}
\includegraphics[width=0.2\textwidth]{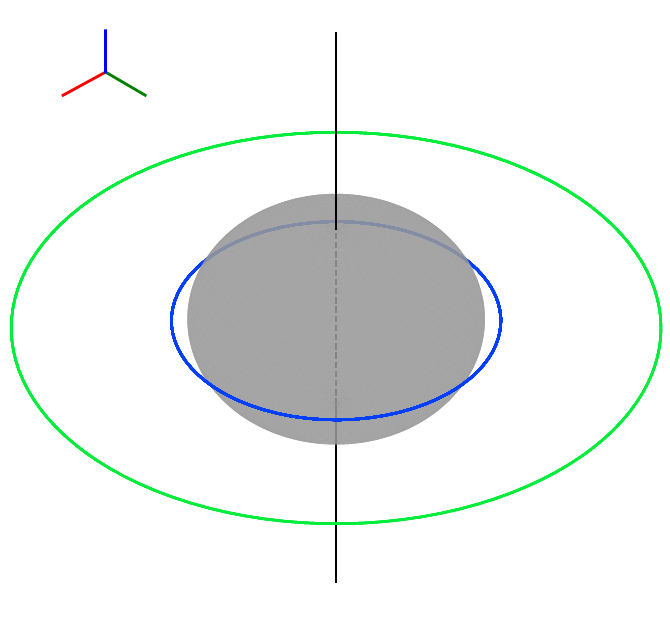} 
\end{tabular}
}
\subfloat[Bound orbit trajectories]
{%
\raisebox{-0.3em}{\includegraphics[width=0.5\textwidth]{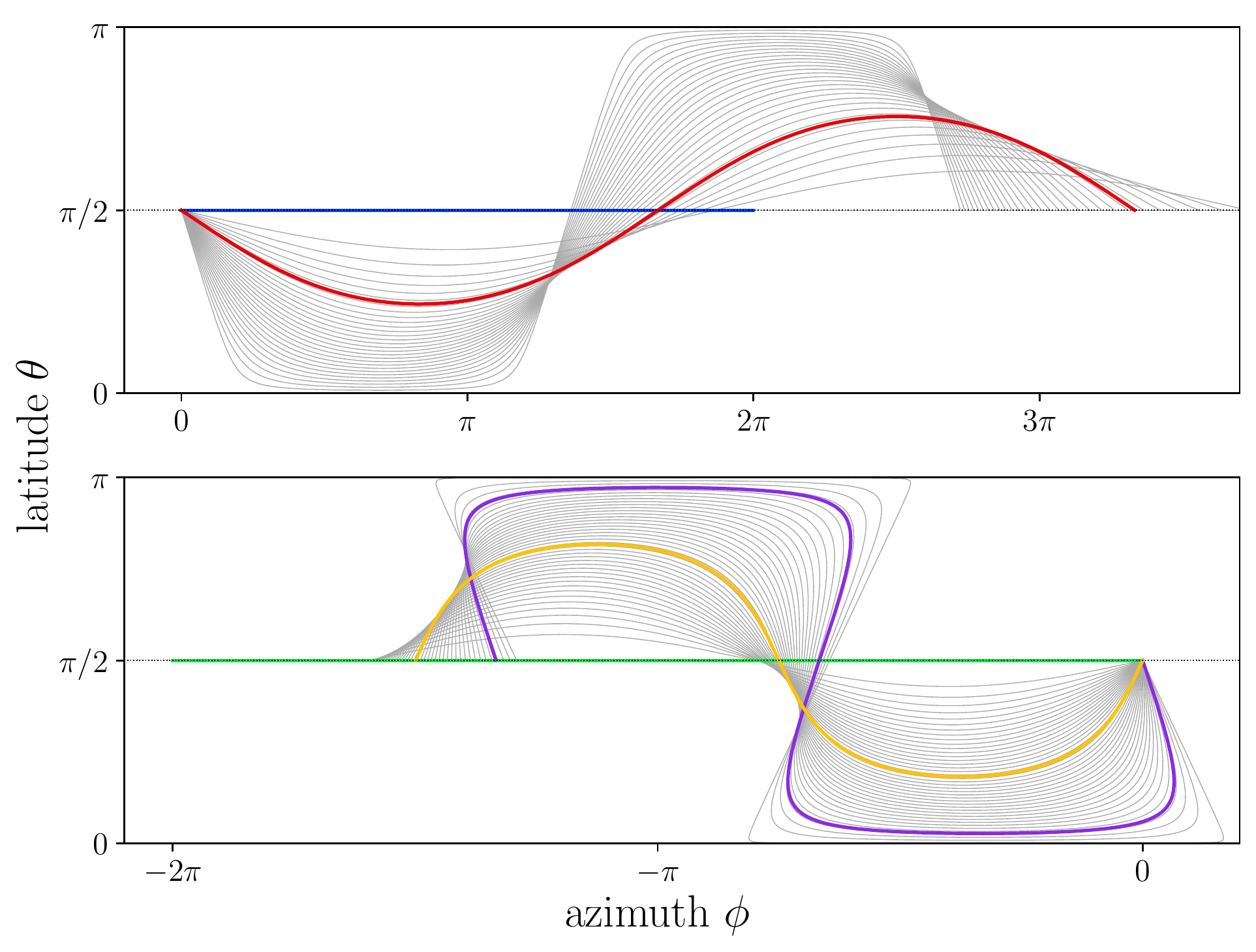}}
}
\end{center}
\caption{
Left panel: Five selected resonant orbits for a spin $\bhspin=4/5$ hole. Right panel: Trajectory of bound orbits for the same $\bhspin=4/5$ hole plotted in the $\theta$--$\phi$ plane. The resonant orbits in the left panel are colored in the right panel.
$\Delta \phi$ for each orbit corresponds to the $\phi$ displacement after one complete latitudinal cycle, i.e., when the trajectory ends on the plot. The spread in $\Delta \phi$ across the different orbits increases with the spin $\bhspin$.
}
\label{fig:orbits3d2d}
\end{figure*}

For holes with nonzero spin, the latitudinal oscillation period is different than the azimuthal $\phi$ period, so after a full $\phi$ orbit around the hole ($\phi \to \phi + 2\pi$), the geodesic will not return to the same $\theta$ coordinate.
The magnitude of the precession can be written in terms of the deviation in $\phi$ from one latitudinal cycle to the next. In the case of no precession, $\Delta \phi$ would be $2\pi$. In general, it is given by
\begin{align}
\label{eq:Deltaphi}
    \Delta \phi &= 4\int\limits_{0}^{\theta_+} \dfrac{\ud \phi}{\ud \theta} \, \ud \theta \nonumber \\
    &= \dfrac{4}{\sqrt{-u_-^2}} \left( \dfrac{2 r - \bhspin \Phi}{\Delta} K\left( \dfrac{u_+^2}{u_-^2}\right) + \dfrac{\Phi}{\bhspin} \Pi \left( u_+^2, \dfrac{u_+^2}{u_-^2} \right) \right),
\end{align}
where $K$ and $\Pi$ are complete elliptic integrals\footnote{We use the square of the elliptic modulus as the parameter for all elliptic integrals in this work.} of the first and third kinds, the roots of the latitudinal potential are
\begin{align}
    \label{eq:upmsqdefn}
    u_{\pm}^2 &\equiv \dfrac{\bhspin^2 - Q - \Phi^2 \pm \sqrt{\left(Q + \Phi^2 - \bhspin^2\right)^2 + 4 \bhspin^2 Q}}{2 \bhspin^2},
\end{align}
and the constants of motion $\Phi(r)$ and $Q(r)$ are given by
\begin{align}
    \Phi &= - \dfrac{ r^3 - 3 r^2 + \bhspin^2 r + \bhspin^2 }{ \bhspin \left( r- 1\right)} \label{eq:Phidef1} \\
    Q &= - \dfrac{r^3\left(r^3 - 6 r^2 + 9 r - 4 \bhspin^2\right) }{\bhspin^2 \left( r - 1\right)^2} \label{eq:Qdef1} .
\end{align}
Figure~\ref{fig:orbits3d2d} plots the trajectory of bound orbits in the $\theta$--$\phi$ plane and illustrates this precession effect, where one complete latitudinal cycle does not correspond to an azimuthal displacement of $2\pi$. The spread in $\Delta \phi$ increases with $\bhspin$.

By studying $\Delta \phi$, we can identify which orbits are also closed---which geodesics return to their original positions and orientations.
Closed (resonant) orbits correspond to the case that $\Delta \phi /2 \pi$ is a rational number $=p/q$ in simplest form. 
Infinitely many closed orbits exist, but orbits with small $p$, $q$ are the most interesting here since they have the shortest path lengths.
Figure~\ref{fig:orbits3d2d} shows the trajectories of several closed orbits both in three dimensions and as projected on the latitude--azimuth plane.

We also compute the time delay for one complete latitudinal cycle of the orbit:
\begin{align}
    \label{eq:timedelaychi}
    \Delta t =& \, 4\int\limits_{0}^{\theta_+} \dfrac{\ud t}{\ud \theta} \, \ud \theta \nonumber \\
    =& \, \dfrac{\left(r^2 + \bhspin^2\right)^2 - 2 \bhspin \Phi r - \bhspin^2 \Delta }{ \bhspin \Delta \sqrt{-u_-^2}} K\left( \dfrac{u_+^2}{u_-^2} \right) \nonumber \\
    & \, - 4 \bhspin \sqrt{- u_-^2 } \left[ K \left( \dfrac{u_+^2}{u_-^2} \right) - E \left( \dfrac{u_+^2}{u_-^2} \right) \right],
\end{align}
where $E$ is the complete elliptic integral of the second kind.
In the case of closed orbits, the time delay to complete a full cycle is a function of the number of latitudinal oscillations per complete cycle:
\begin{align}
    \Delta T = q \; \Delta t \; 2 \pi / \Delta \phi .
\end{align}

\pagebreak

\section{The image of the photon shell} \label{sec:photonshellobs}

We now consider the signature of the photon shell as seen by an observer far away from the hole.
Since the orbits comprising the shell are unstable, photons whose paths deviate slightly from the precise trajectories described above will either fall onto the hole or escape to infinity where they can be captured by an observer.
The exponential instability can be described in terms of how the deviation $\delta r$ between a geodesic's radial position and the radial position of the corresponding bound orbit increases (or decreases) after $n_\phi$ azimuthal cycles around the hole
\begin{align}
    \delta r(n_\phi) = \exp \left( \pm \gamma_\phi n_\phi \right) \delta r(0) .
\end{align}
This equation defines the Lyapunov exponent $\gamma_\phi$, which is given by
\begin{align}
    \gamma_\phi = \dfrac{2\pi}{\Delta \phi} \; 2 \gamma_\theta
\end{align}
where $\gamma_\theta$ is taken to be consistent with \cite{johnson2020universal} and governs the deviation after one latitudinal $\theta$ half-cycle (from midplane to extremum back to midplane):
\begin{align}
    \gamma_\theta 
    &= \dfrac{4}{\bhspin \sqrt{-u_-^2}} \sqrt{ r^2 - \dfrac{ r \Delta}{\left(r-1\right)^2} } K\left( \dfrac{u_+^2}{u_-^2} \right) \; .
\end{align}
This Lyapunov exponent, the time delay of Equation~\ref{eq:Deltaphi}, and the azimuthal period of Equation~\ref{eq:timedelaychi} characterize strong lensing by Kerr black holes \citep{gralla2020criticalexp}.

Rather than consider the source-to-observer model, in which we start with all geodesics that are emitted from a source and select only those that make it to the observer, it is convenient to switch to the observer-to-source model, where we begin with the set of all geodesics that intersect the image plane.
For a camera at infinite distance, all photons incident on the camera arrive parallel to each other. 

\begin{figure}[ht!]
\begin{center}
\includegraphics[width=\linewidth]{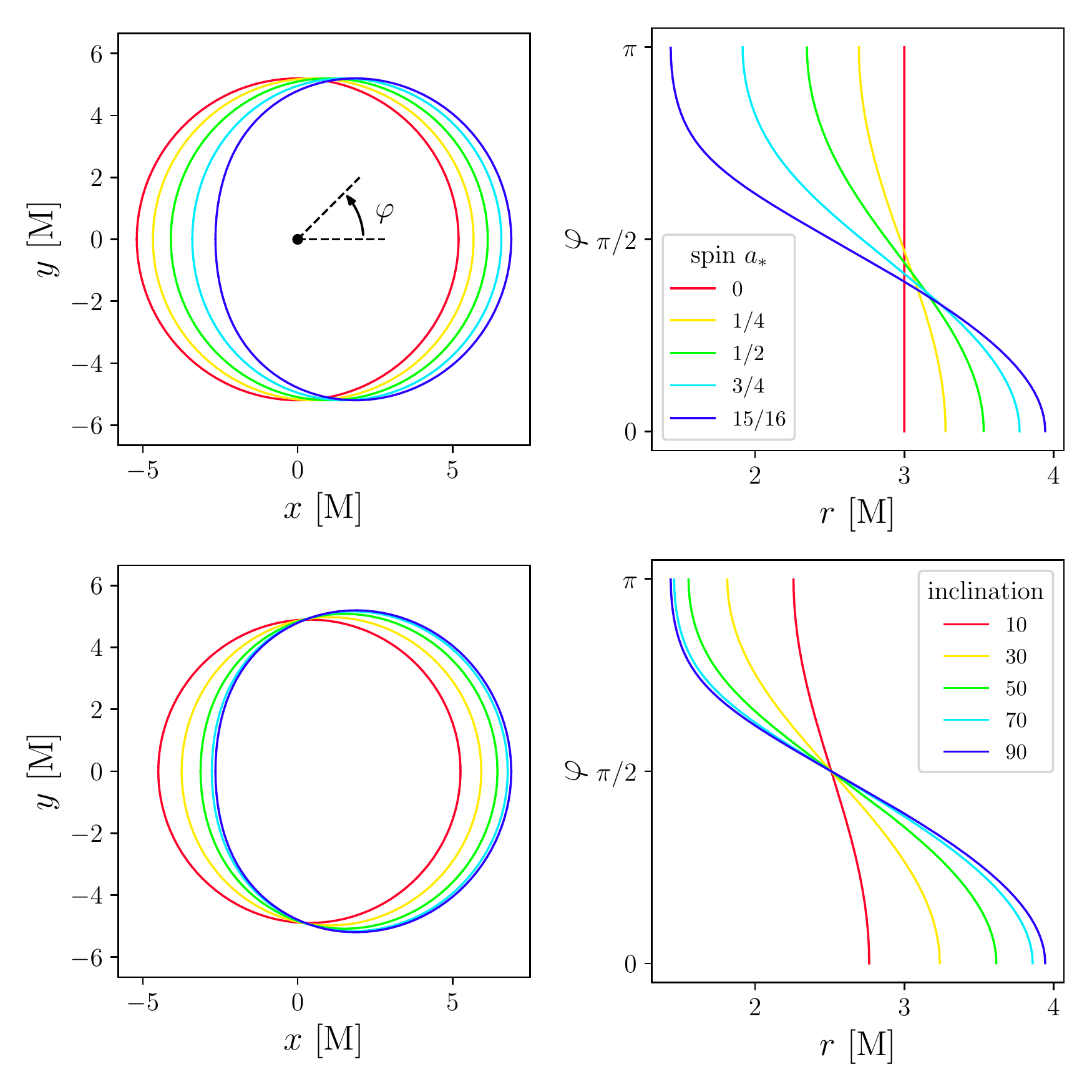} 
\end{center}
\caption{
Left column: size and shape of critical curve on image for (top) black holes with different spin viewed edge-on or (bottom) black holes with $\bhspin=15/16$ viewed at different inclinations. The top-level panel shows how the angle $\varphi$ is measured counterclockwise from the positive $x$-axis. Right column: mapping between points along the critical curve $\varphi$ and the Boyer-Lindquist radius of the probed bound orbit.
}
\label{fig:criticalcurvemapping}
\end{figure}

Following \cite{bardeen1973kerrshadow}, we parameterize geodesics that intersect the image plane according to their impact parameters $x$ and $y$ ($\alpha$ and $\beta$ in Bardeen). By convention, the $y$ axis is aligned with the projection of the black hole spin axis on the image. It is sometimes more convenient to work with polar coordinates on the image $\rho = \sqrt{x^2 + y^2}$ and $\varphi = \arctan y / x$ (see the top left panel of Figure~\ref{fig:criticalcurvemapping}).

Geodesics at large impact parameter (far from the image center) remain far from the hole and barely feel its influence. As the impact parameter decreases, however, the geodesics become increasingly bent, and eventually they wrap around the hole and undergo latitudinal oscillations. The set of impact parameters for all geodesics that undergo a fixed number of oscillations defines a subring on the image.

Nested subrings produce demagnified images of space. The magnification is given by the image area of the subring and scales according to the Lyapunov treatment above. If most of the emission is produced near the black hole, then it will be sampled $n$ times in the $n$th subring, so the ratio of flux (area-integrated intensity over the subring on the image) between the $n$ and $n+1$ subrings will be given by
\begin{align}
    \label{eq:lyapunovflux}
   \dfrac{F_{n}}{F_{n+1}} \approx \dfrac{ e^{-\gamma(n-1)} - e^{-\gamma n}}{e^{-\gamma n} - e^{-\gamma (n+1)}} = e^{\gamma} .
\end{align}

In the limit that $n$ goes to infinity, geodesics wrap around the hole infinitely many times and are effectively trapped by the bound orbits described above. The set of all impact parameters with $n \to \infty$ defines the \emph{critical curve} on the image. The region within the critical curve is sometimes called the black hole shadow. 

Points on the critical curve correspond to different bound orbits---different radii---within the photon shell, so the path of the curve can be parameterized by $r$ (see  \citealt{bardeen1973kerrshadow} and also \citealt{Johannsen2013}):
\begin{align}
    x &= - \dfrac{\Phi}{\sin i} \\
    y &= \pm \sqrt{ Q + \bhspin^2 \cos^2 i - \Phi^2 \cot^2 i} .
\end{align}
The shape of the critical curve is thus a function of two parameters: the spin of the hole and the viewing inclination angle $i$.
Figure~\ref{fig:criticalcurvemapping} shows the shape of the critical curve both for black holes with different spins (top) and for the same hole observed at different inclinations (bottom). It also shows the mapping between points along the critical curve (parameterized by $\varphi$) and the bound orbits they correspond to (parameterized by $r$). The orientation of the curve is determined by the projection of the black hole spin axis on the image plane.

The spin of the hole determines which radii support bound orbits.
Varying the inclination angle limits the set of bound orbits that are visible to the observer (for example, purely equatorial orbits are not accessible to a top-down observer). The set of accessible radii decreases from the full range given by Equaion~\ref{eq:rminmax} at $i=90^\circ$ to the the single radius corresponding to $\Phi = 0$ at $i=0^\circ$ (for more detail, see Appendix~\ref{sec:appendix:kerreom} and especially Equation \ref{eq:appendix:Phi0}).

\section{The Kerr Glimmer Signature} \label{sec:echospectrum}

Although messy gas dynamics determine the larger image features, the photon shell produces a separable, unique signature in the image domain. This signature is independent of the gas dynamics model, so it provides a direct way to measure the properties of the underlying black hole.
The presence of an infinite number of subrings on the image means that an (optically thin) emission source will be imaged an infinite number of times, albeit with exponentially decreasing flux from one instance to the next. The image in the $n$th subring comes from light that has gone around the hole $n$ times, and each subring images will echo with a period equal to the total light travel time around the hole.

The aggregate signature produced by an emission source depends on the characteristics of the source, but we can provide an initial analysis by making two remarks in the context of a simplified model. First, since different positions along the critical curve correspond to different bound orbits and path lengths, the delay between subsequent imagings will be a function of position on the curve.

Second, if the source emission is localized in space, then an orbit that probes the source must return to the same localized area in order for an echo to be excited along that geodesic. Since bound orbits precess, a geodesic that passes through a fiducial source may take many revolutions around the hole before it passes through the source again.
Since flux decreases exponentially with $n_\phi$, the precession can render some orbits practically echoless. The latter criterion is the most restrictive and is relaxed in the second model we present.

In practice, the echo response function to an individual emission event is complicated since it depends on the size and duration of the source. The details are complicated further by the initial transient response, which is determined by source size, duration, and position.
Rather than attempt a full treatment of the emission variability near the hole, we consider simplified hotspot models. In Section~\ref{sec:discussion}, we argue that the hotspot model is not restrictive, especially in the limit of an optically thin source.

Each hotspot is taken to be a transient, isolated Gaussian blob with a time- and position-dependent emissivity
\begin{align}
    j(t, \vec{r}) &\propto \Psi(t) \exp \left( -\delta r^2 / 2 \sigma_r^2 \right) \exp \left( -\delta t^2 / 2 \sigma_t^2 \right),
\end{align}
where $\delta r$ is the distance to a point on a Keplerian orbit at radius $r_0$, $t_0 \equiv t - \delta t$ sets the time when the hotspot is brightest, $\sigma_r$ and $\sigma_t$ describe the width of the hotspot in space and time, and
where $\Psi(t)$ is a bump function in time
\begin{align}
\Psi(t) &= 
    \begin{cases} 
      \exp \left( \dfrac{1}{ \delta t^2 - \sigma_t^2 } \right) & \left| \delta t \right| < \sigma_t \\
      0 & \mathrm{otherwise},
   \end{cases}
\end{align}
which forces a smooth decay to zero emissivity.

\begin{figure}[ht!]
\begin{center}
\includegraphics[width=\linewidth]{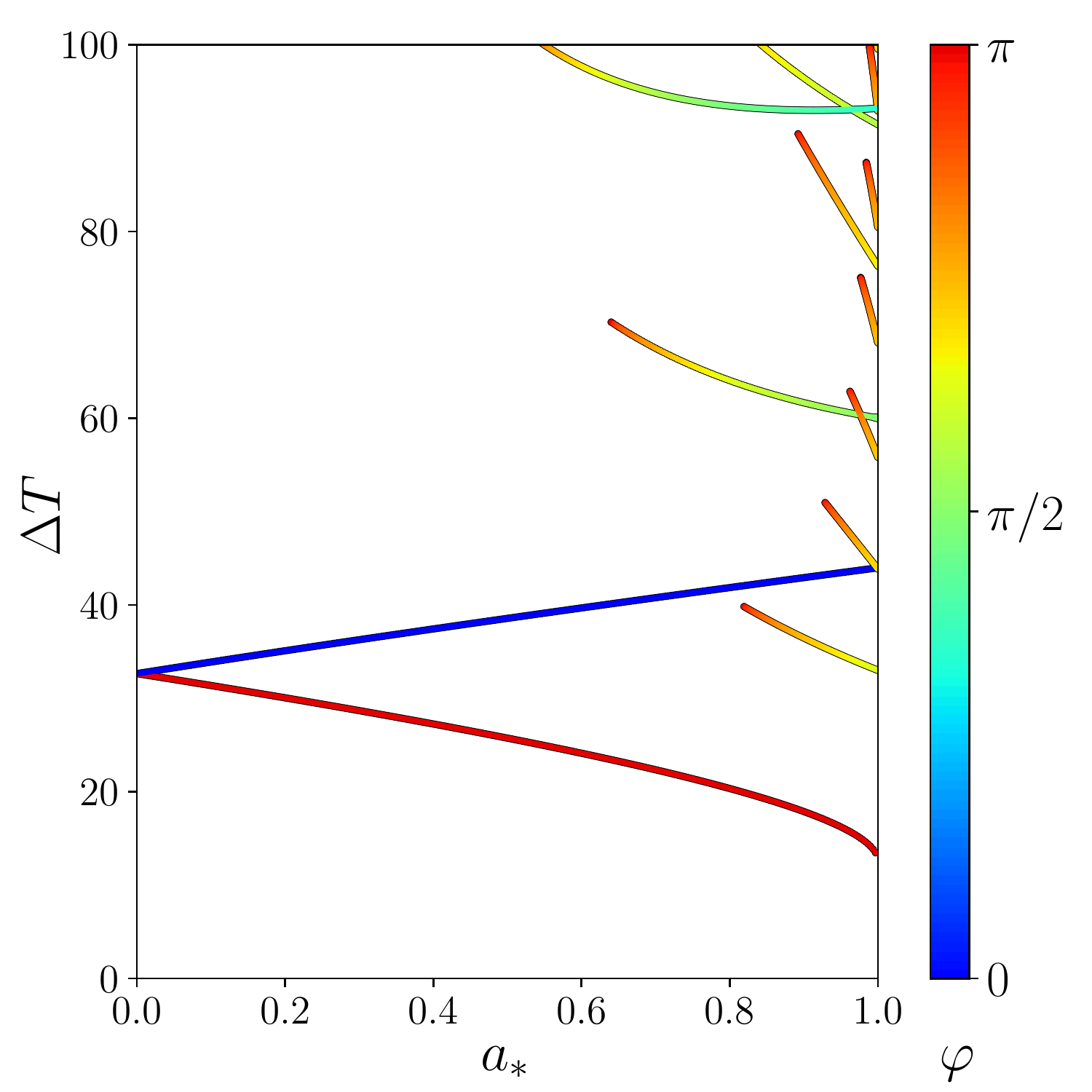} 
\end{center}
\caption{
Echo delay times due to resonant orbits as a function of black hole spin. As spin increases, the number of accessible short-time-delay echoes increases. Color encodes the location of the echo on the critical curve. Only perfect resonances are shown in this plot. 
}
\label{fig:delayspectrum}
\end{figure}

\subsection{Point source emission}

\begin{figure*}[ht!]
\begin{center}
\includegraphics[width=0.96\linewidth]{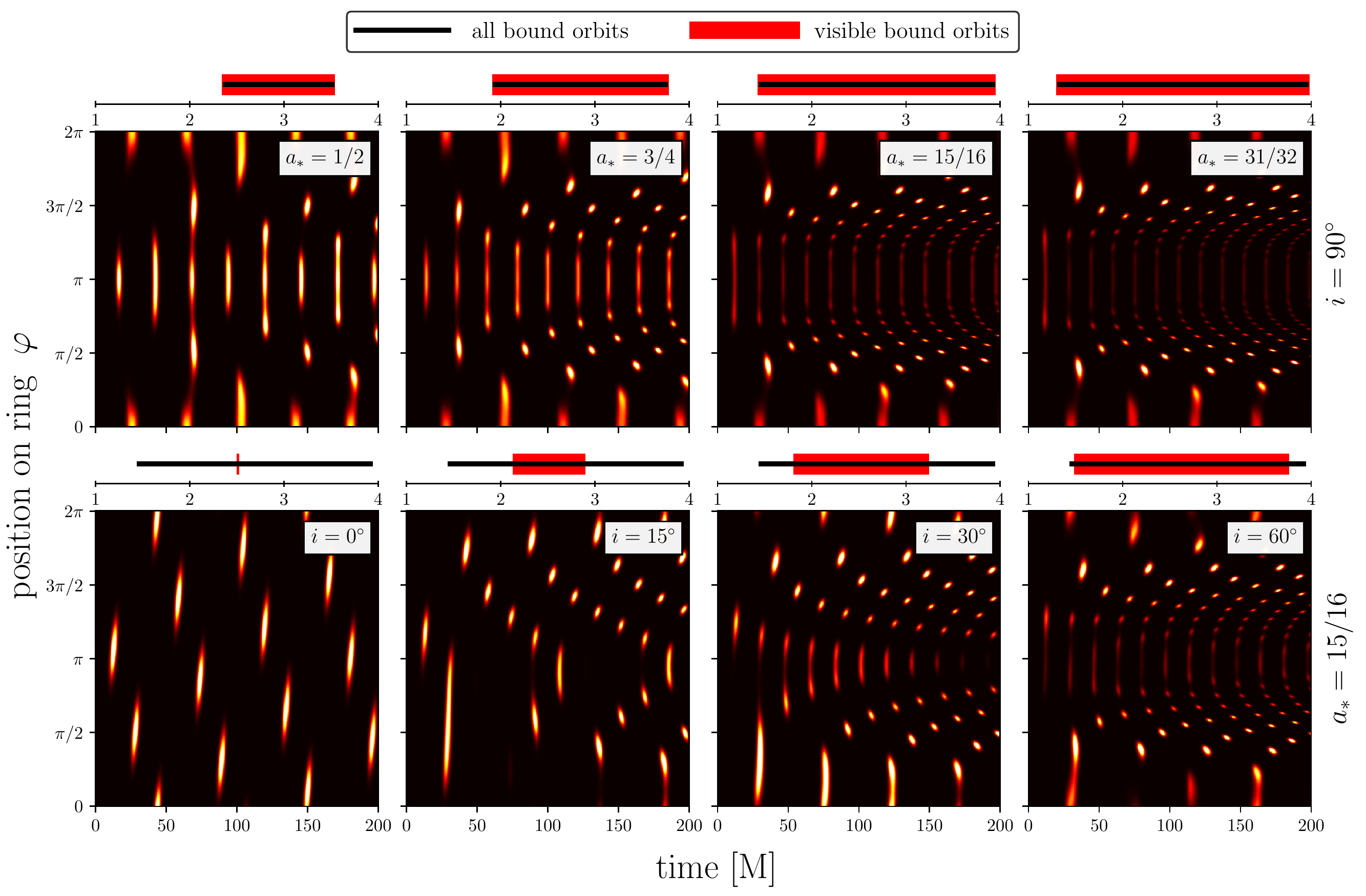}
\end{center}
\caption{
Echo response produced by a short-lived, Keplerian hotspot ($r_0=2.8\,\mathrm{M}$, $\sigma_r=0.5\,\mathrm{M}$, $\sigma_t=1\,\mathrm{M}$) for black holes with different spins and for observers at different inclinations. Color encodes the time-dependent intensity of light as a function of angle along the critical curve $\varphi \in (0, 2\pi)$. The bar above each panel shows the range of radii corresponding to bound orbits (black) and the set of visible bound orbits (red). Top row: the variation in the response as a function of black hole spin. The spread in echo delays between the prograde orbit ($\varphi = \pi$) and the retrograde orbit ($\varphi = 0, 2\pi$) increases as spin is increased. Bottom row: the echo response for the same $\bhspin = 15/16$ hole but observed at different inclinations. As inclination increases, the set of visible bound orbits increases, so more echoes become visible. Since the $\Phi=0$ orbit is not closed for $\bhspin = 15/16$, the $i=0^\circ$ echo blips drift through $\varphi$. An animation showing the evolution of the echo response as a function of spin and inclination is available online.
}
\label{fig:spinincglimmer}
\end{figure*}

We start by considering the response produced by a unidirectional point source emitter as measured along the critical curve. Since echoes occur when geodesics sample the same emission source multiple times, then in the limit as $\sigma_r \to 0$, the only universal echoes that occur at a fixed $\varphi$ must come from closed bound orbits, since closed orbits are the only ones that return to the exact same point while following the exact same heading.
Other (non-closed) geodesics will either miss the emitter, not arrive at the correct observer location, or pass through the emitter with a different heading and thus accrue no new intensity. The potential for an arbitrary emission source to be imaged but not produce a trivial echo is what differentiates Kerr from Schwarzschild, since all spherical orbits in Schwarzschild are closed.

Even though an infinite number of closed orbits exist, only orbits with short path lengths will produce observable echoes since the sensitivity required to detect echoes from multiple turnings increases exponentially with the number of turnings according to the Lyapunov treatment summarized by Equation~\ref{eq:lyapunovflux}. Thus for a given spin, there is a limited number of fixed, position-dependent observable echo time delays. Notice that although the signal from echoes with longer delays is more attenuated, the overall magnitude of the dimming may decrease as spin is increased.

Figure~\ref{fig:delayspectrum} plots the allowed echo delays due to closed orbits as a function of spin. Because each delay is due to a particular bound orbit, it maps to a distinct position along the critical curve. 
The allowed delays in the figure are colored according to the angle  $\varphi$ along the critical curve where they appear. The time delays for the prograde and retrograde orbits are always accessible since they lie within the equatorial plane and thus always pass through the same points. The closed orbits can be computed by identifying which radii in Equation~\ref{eq:Deltaphi} correspond to (the reciprocals of) the first few levels of a Stern--Brocot tree. The non-trivial radial structure of the Kerr spacetime means that the time delay for a geodesic that circles the hole twice before closing will not be twice the delay for a geodesic that only circles it once, so the higher-order resonances are not perfect multiples of the lower-order ones. 

Thus far, we have considered unidirectional emitters. If this restriction is removed, new echoes with different periods may be produced when a geodesic intersects itself along a different heading. For example, closed orbits with an even-to-odd ratio between the number of azimuthal and latitudinal cycles will return to the same point (with mirrored latitudinal heading) after having completed only half of the full closed path. Echoes due to self-intersections off of the midplane (see Figure~\ref{fig:orbits3d2d} for examples) will lead to more complicated, composite delay structures that repeat with the same fundamental periodicity of the closed orbits. We return to this point in our discussion of hotspot position.

\begin{figure}[ht!]
\begin{center}
\includegraphics[width=0.98\linewidth]{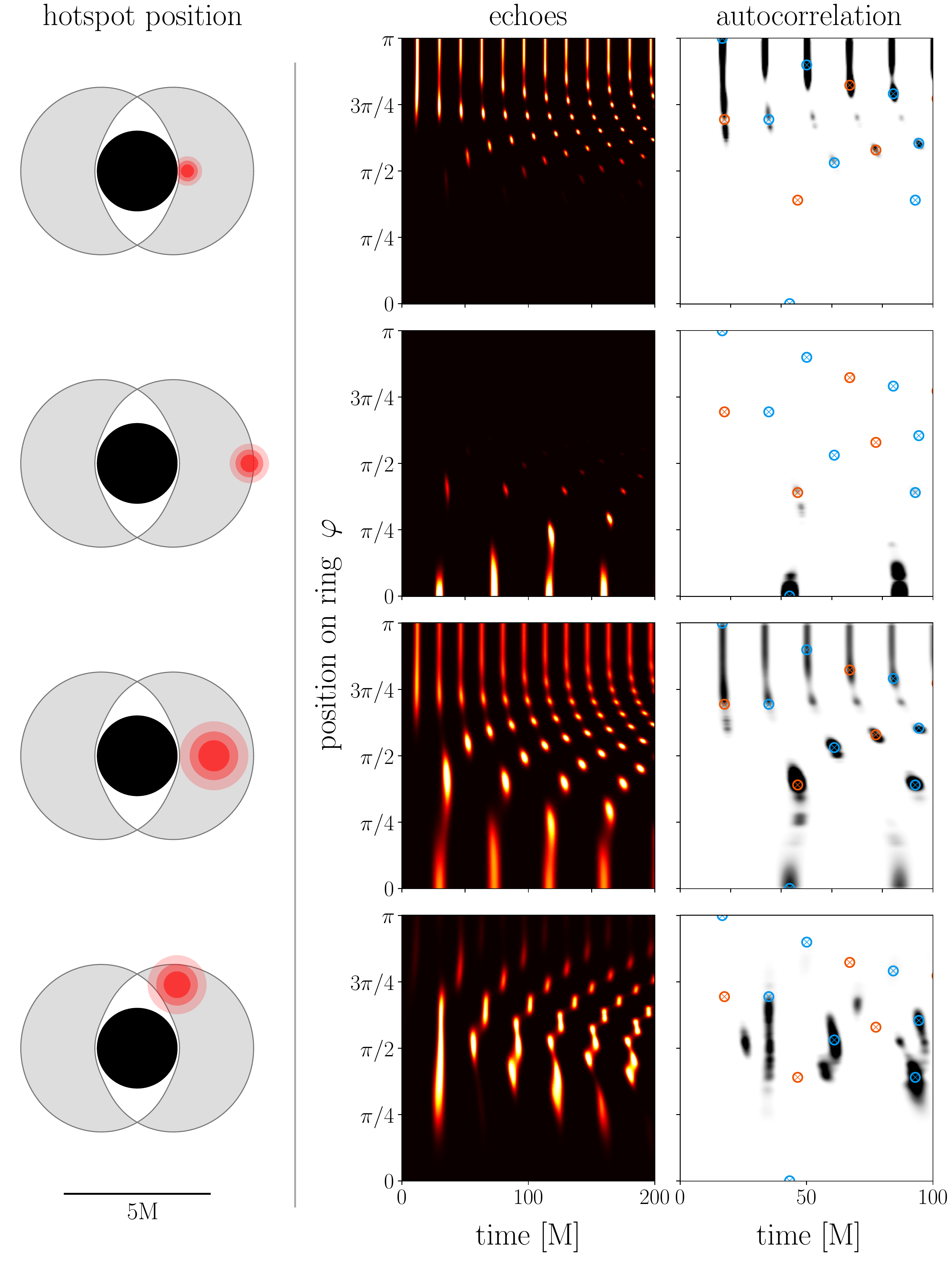}
\end{center}
\caption{
Effect of hotspot position and size on echo response. Left column: position and size of hotspot (on Keplerian orbit with $\sigma_t = 1\,\mathrm{M}$) in space around a black hole (black circle) with $\bhspin=15/16$. The gray shaded crescents mark the region containing the bound orbits. Central column: echoes produced by the hotspot (as in Figure~\ref{fig:spinincglimmer}). Right column: autocorrelation of echoes. Blue $\otimes$'s denote pure glimmer echoes, and orange $\otimes$'s correspond to midplane echoes. Midplane echoes are excited in the first three rows.
When the hotspot is raised above the midplane in the bottom panel, the supplemental echoes no longer peak at \textonehalf~the fundamental glimmer period. In contrast, the glimmer echoes are universal: if a given $\varphi$ exhibits any echoes, it will exhibit echoes at the blue $\otimes$'s. Since the hotspot in the fourth panel does not intersect the (equatorial) prograde and retrograde orbits, it does not excite echoes along those $\varphi$.
An animation showing how the echoes and autocorrelation change as a function of hotspot size and position is available online.
}
\label{fig:hotspotposition}
\end{figure}

\subsection{Finite-width sources}
\label{sec:finitewidth}

Real world emission sources have non-zero width. In the context of glimmer, increasing the width of an emission sources softens the condition that a geodesic must return to the same point in ($\theta, \phi$) space for an echo to be produced, since geodesics will resample the fiducial source feature as long as the deviations in their positions are smaller than the size of the feature. These more permissive conditions broaden the set of accessible time delays and allow a wider range of $\varphi$ to clearly echo.

The echo response can be visualized by plotting the intensity of light observed at different angular positions along the critical curve as a function of time. Black holes with different spins and observers at different inclinations will see echoes produced by bound orbits at different radii. Figure~\ref{fig:spinincglimmer} shows how changing the spin or inclination affects the echo response produced by a Keplerian hotspot with $r_0=2.8\,\mathrm{M}$, $\sigma_r=0.5\,\mathrm{M}$, and $\sigma_t = 1\,\mathrm{M}$. For each echo response, the radii that admit bound orbits (determined by spin) is identified by a black line, and the subset of radii with visible bound orbits (determined by inclination) is denoted by a red bar.

The top row of Figure~\ref{fig:spinincglimmer} shows how the echo response changes as the angular momentum of the central black hole increases. As seen in Figure~\ref{fig:delayspectrum}, the difference between the echo delays for the prograde ($\varphi = \pi$) and retrograde ($\varphi = 0, 2\pi)$ orbits increases as spin increases.
The bottom row of Figure~\ref{fig:spinincglimmer} shows the effect of varying inclination for a black hole with fixed spin $\bhspin = 15/16$. At the top-down $i=0^\circ$ inclination, every point along the critical curve corresponds to an orbit at the same $\Phi=0$ radius (see Equation~\ref{eq:appendix:Phi0}), so the echo delay period is constant around the curve. Notice that for $\bhspin = 15/16$, the $\Phi = 0$ orbit is not closed, so the echo blips drift around the curve. As inclination increases to $i=90^\circ$, the full set of bound orbits becomes visible.

\vspace{1em}

The glimmer signature comprises a mapping between echo delay periods and positions along the critical curve.
Since this mapping is due solely to the spacetime geometry, it is independent of emission details like hotspot position, shape, and size. In contrast, the details of the full echo response, such as the relative strengths or phase offsets of echoes at different $\varphi$, are influenced by the characteristics of the hotspot.

Figure~\ref{fig:hotspotposition} shows the echo responses produced by different hotspots. In addition to plotting the echo response (as in Figure~\ref{fig:spinincglimmer}), Figure~\ref{fig:hotspotposition} also shows the autocorrelation of the measured signal as a function of $\varphi$. 
Since the autocorrelation function computes relative delays, it intrinsically separates the echo periods from phase offsets around the ring. The glimmer signature naturally lives in this autocorrelation space; the glimmer mapping is plotted in the figure as a set of blue $\otimes$'s. Half-period echoes, which can be excited by emission in the midplane, are marked as orange $\otimes$'s.

The subset of echoes that are excited is determined by the set of bound orbits that the source intersects. As the source size increases, the $\varphi$ width of each response blip increases as nearly closed orbits begin to resample the source. 
When the hotspot is in the midplane, it excites the half-period echoes. 
As the hotspot moves away from the midplane, the autocorrelation response drifts away from the orange $\otimes$'s. Also, as the hotspot moves from the midplane to low inclination, it intersects fewer bound orbits (e.g., consider the fact that the prograde and retrograde orbits lie entirely within the midplane). Thus, in addition to producing responses with more complicated delay structures, hotspots off the midplane cannot excite the full spectrum of glimmer echoes.

While the echoes described by glimmer are always excited, geodesic self-intersection excites supplemental echoes based on source features. If the majority of the flux in an observation is produced by a single, transient hotspot, then the additional echo structure may provide a means to infer details about the hotspot. In contrast, if the source emission is extended, diffuse, and stochastic, the supplemental echoes will be washed out by the universal glimmer signature.

\section{Discussion} \label{sec:discussion}

We have described black hole glimmer, a position- and time-dependent effect in black hole images that is due to bound and closed photon orbits.
The set of echo periods and arrival locations that comprise glimmer uniquely encodes the properties of the spacetime and provides a direct way to measure black hole mass and angular momentum.
Since the universal glimmer signature is determined only by the underlying geometry, it is separable from astrophysical and plasma uncertainties in the emission model. 
We now briefly present a simulated glimmer measurement at finite resolution, discuss the generality of the hotspot model, and consider extensions to our analysis.

\subsection{Measuring glimmer with finite resolution}

Recently, \citet{Hadar2020autocorrelations} discussed the potential of performing a measurement of autocorrelations for a nearly top-down black hole with a thin surface emissivity in the midplane, and \citet{Chesler2020} treated the potential of performing a measurement of coherent autocorrelations in a similar context.
The clearest glimmer signatures are encoded in the time- and position-dependent intensities along the critical curve, but performing a measurement with sufficient resolution to resolve the detailed spatial dependence may be technologically impractical.
Although evaluating the feasibility of a detailed measurement of glimmer is beyond the scope of this paper, we now perform a simple example glimmer measurement from a simulated low-resolution observation.

\begin{figure}[ht!]
\begin{center}
\includegraphics[width=\linewidth]{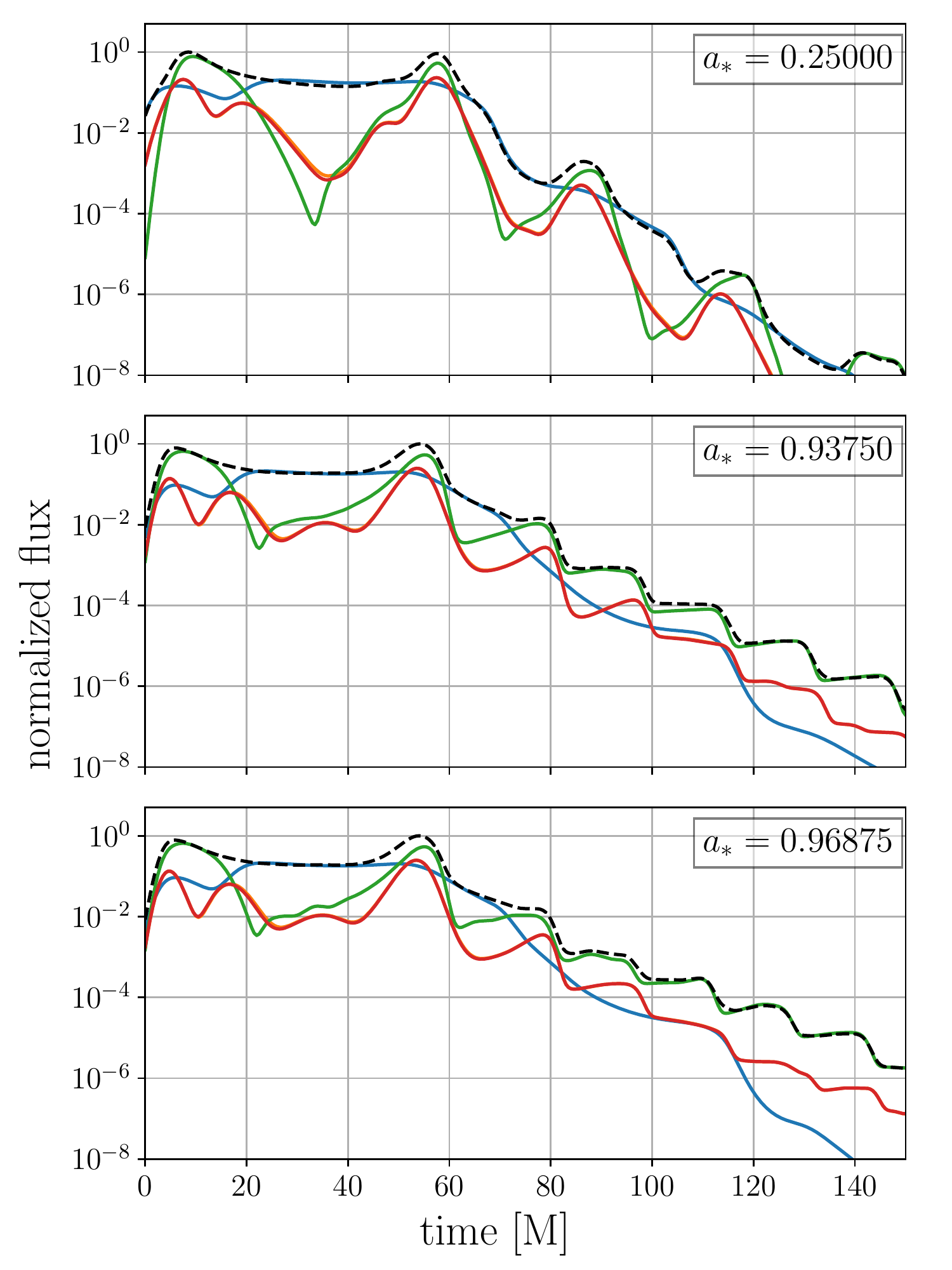} 
\end{center}
\caption{Quadrant-based light curve decomposition for a hotspot at $r_0 = 3$ M and $r_\sigma = 0.8$ M with a flat emissivity profile versus frequency orbiting around black holes with $\bhspin = 1/2, 15/16$, and $31/32$. The full light curve (black dashed line) is divided into four image quadrants: retrograde-centered $\varphi\in(\pi/4, -\pi/4)$ (blue), prograde-centered (green) $\varphi \in (3\pi/4, 5\pi/4)$, and two remaining (red) regions. The prograde orbit echoes are strongest because they have the smallest Lyapunov exponents. The initial transient to $t \approx 60$ GM/c$^3$ is produced during the time that the hotspot is active and orbiting the hole. 
}
\label{fig:splitlc}
\end{figure}

Even with limited resolution, the echo structure can be probed by comparing the light curves produced by different regions in an image. Figure~\ref{fig:splitlc} shows the light curves measured in different regions of an angular image decomposition as seen by an observer at $i=90^\circ$. The different panels show the decomposition for a numerically simulated hotspot ($r_0 = 3\,\mathrm{M}$, $\sigma_r = 0.8\,\mathrm{M}$, and $\sigma_T = 20\,\mathrm{M}$) on a Keplerian orbit in the midplane around three different black holes with spins $\bhspin = 1/4,\ 15/16$, and $31/32$. 
After the initial transient decays, the unresolved (total) light curve signal is dominated by the prograde echo period, but since the different light curves correspond to different parts of the image---and thus cover different arcs along the critical curve---they echo with different periods.

As Figure~\ref{fig:splitlc} shows, it may be possible to resolve different echo periods among the different light curves. Even if the precise time delay between peaks in a single light curve is not measurable, the spread in delays could be used to bound the spin and mass of the system. A rough comparison of the delay periods could also be performed to bound the orientation of the curve and infer the orientation of the spin axis of the hole on the sky.

\subsection{Hotspot model generality}

We evaluated the glimmer signature produced by a transient hotspot in a Keplerian orbit around a black hole; however, this toy emission model may not be representative of typical astrophysical scenarios. Real world emitters likely follow more complicated trajectories, and changing the velocity of the emitting material can contribute to effective source anisotropy via relativistic beaming effects. As the relative amplitude of the signals emitted in different directions changes, the relative amplitude of echo signatures at different $\varphi$ will change as well. Additionally, real emission sources may comprise diffuse structures that are not well-approximated by localized hotspot-like emission.

Nevertheless, our model approximation is not restrictive because any realistic source can be decomposed into a set of localized, transient emission features. In the limit of zero optical depth, the echo signatures produced by different hotspot add linearly, with each component weighted by the relative strength of the underlying emission feature. Thus, although the emission structure can influence relative intensity of the echoes, the position-dependent delay spectrum is universal.

If the emission source is localized, then it may be possible to constrain the vertical location of the emission by subtracting the glimmer echoes from the full observed echo (autocorrelation) spectrum. We note, however, that the transient signatures produced by direct emission have much higher flux relative to the echoes and are therefore likely to provide a much more efficient means of inferring source structure.

It may also be possible to measure glimmer produced by an emitter far from the hole, e.g., especially in the case of a stellar mass black hole. If a coherent light source is located directly behind the hole relative to the observer, then the light paths between the source and the observer will undergo the same lensing effects as they pass the hole.

\subsection{Limitations and extensions}

In our treatment, we assumed that the emission was independent of frequency; in sources with non-trivial frequency dependence, redshift effects can change the intensity of light received at different $\varphi$ around the critical curve and thus may influence the sensitivity required for a measurement.
If the emission spectrum has a characteristic frequency, then it may be possible decompose an unresolved light curve and measure glimmer by comparing the dominant echo periods of different components of the spectrum. Since each bound orbit follows a different latitudinal profile, different orbits can intersect the source at different angles relative to the source motion. Frequency shift is controlled by this angle, so different orbits (and thus different echo periods) will peak with different characteristic frequencies. We also neglected the effect of optical depth, which decreases the strength of high order echoes compared to the analytic Lyapunov treatment. Since optical depth is a function of wavelength, it may be desirable to observe at frequencies where the optical depth of the plasma is minimal.

More broadly, our treatment neglected the initial transient (which is a strong function of the position, shape, and dynamics of the source).
An analysis of general relativistic magnetohydrodynamic simulations  may be required to study the detailed complexities of a true observation; an analytic treatment of correlations must faithfully reproduce the spatial structure of the emission source, since source position determines which echoes are excited.

\section{Summary} \label{sec:summary}

We have described black hole glimmer, a position- and time-dependent echoing in black hole images that is due to bound and closed photon orbits.
The set of echo periods and arrival locations that comprise glimmer uniquely encodes the properties of the spacetime and provides a robust, independent probe of black hole mass and angular momentum. Glimmer makes precise predictions that could be used to directly test the Kerr hypothesis.
Since glimmer is determined only by the underlying geometry, its signature is separable from astrophysical and plasma uncertainties in the emission model. 
We have used numerical simulations to demonstrate that the glimmer signature may be observable even in a limited-resolution measurement.

\acknowledgements

I am grateful to Avery Broderick, Paul Chesler, Shahar Hadar, Alex Lupsasca, Kotaro Moriyama,  Ramesh Narayan, Nico Yunes, and especially Charles Gammie for clarifying conversations and comments. I also thank the anonymous referee for helpful comments and suggestions that greatly improved the text. This work was supported by NSF grant AST 17-16327 and by a Donald C.~and F.~Shirley Jones Fellowship. 

\appendix

\section{Kerr Equations of Motion}
\label{sec:appendix:kerreom}

The Kerr metric is cyclic in the $t$ and $\phi$ coordinates, so it has two killing fields $\partial_t$ and $\partial_\phi$. These correspond to the conserved conjugate momenta $p_t$ and $p_\phi$ that are canonically associated with (negative) energy at infinity and angular momentum about the spin axis of the hole. 
\citet{Carter1968} identified a third conserved quantity $\qcarter$ that is associated with a second-order killing tensor field and is physically related to the $\theta$ velocity of a particle as it passes through the midplane.

For a test particle with four-momentum $p^\mu$, the three conserved quantities (\ref{eq:qcarter})--(\ref{eq:zangularmomentum}) along with the particle's mass $-\mu^2 = p^\mu p_\mu$ uniquely determine geodesics in the Kerr spacetime.
\begin{align}
    \qcarter &= p_\theta^2 + \cos^2 \theta \left( \bhspin^2 \left( \mu^2 - p_t^2 \right) + p_\phi^2 / \sin^2 \theta \right) \label{eq:qcarter} \\
    \widetilde{E} &\equiv - p_t \label{eq:energy} \\
    L_z &\equiv p_\phi \label{eq:zangularmomentum}
\end{align}

Photons follow null geodesics $x^\mu(\lambda)$, where $\lambda$ is an affine parameter such that $u^\mu = \dot{x}^\mu \equiv \ud x^\mu / \ud \lambda$ and $u^\mu u_\mu = 0$. 
The equations of motion for null geodesics around a black hole are scale invariant with respect to the mass of the hole are can be written as a one-parameter set of ordinary different equations \citep[][]{Carter1968, Bardeen1972}
\begin{align}
    \Delta \Sigma \dot{t} &= \left( \left(r^2 + \bhspin^2\right)^2 - \Delta \bhspin^2 \sin^2\theta\right) \widetilde{E} - 2 r \bhspin L_z \\
    \Sigma^2 \dot{r}^2 &= \widetilde{E}^2r^4 + \left(\bhspin^2 E^2 - L_z^2 - \qcarter\right)r^2 + \nonumber \\
    &\qquad 2 \left(\left( \bhspin \widetilde{E} - L_z\right)^2 + \qcarter \right) r - \bhspin^2 \qcarter \label{eq:rdot} \\
    \Sigma^2 \dot{\theta}^2 &= \qcarter - \left( L_z^2 / \sin^2 \theta - \widetilde{E}^2 \bhspin^2 \right) \cos^2 \theta \label{eq:thetadot} \\
    \Delta \Sigma \dot{\phi} &=  2 r \bhspin \widetilde{E} + \left( \Sigma - 2 r \right) L_z / \sin^2 \theta .
\end{align}

Since photons are massless and their paths are independent of their energies $\widetilde{E}$, it is convenient to normalize both $L_z$ and $\qcarter$ by $\widetilde{E}$ to define new constants of motion $\Phi$ and $Q$. These two constants can be written in terms of the orbit radius $r$ and are often used to parameterize the bound orbits around a hole of a given spin (see the equivalent Equations~\ref{eq:Phidef1} and \ref{eq:Qdef1}).
\begin{align}
    \Phi &\equiv L_z / \widetilde{E} = - \dfrac{ r^3 - 3 r^2 + \bhspin^2 r + \bhspin^2 }{ \bhspin \left( r- 1\right)} \label{eq:appendix:Phidefn} \\
    Q &\equiv \qcarter / \widetilde{E}^2 = - \dfrac{r^3\left(r^3 - 6 r^2 + 9 r - 4 \bhspin^2\right) }{\bhspin^2 \left( r - 1\right)^2} \label{eq:appendix:Qdefn} .
\end{align}

By definition, spherical orbits lie at fixed, unchanging radii, $\dot{r} = \ddot{r} = 0$. By solving Equation~\ref{eq:rdot} for these two conditions and rejecting non-physical solutions (see \citealt{Teo2003} for more detail), we reproduce Equation~\ref{eq:rminmax} and find that spherical orbits must lie between
\begin{align}
    r_\pm = 2 \left( 1 + \cos \left( \dfrac{2}{3} \cos^{-1} \pm \bhspin \right) \right).
    \label{eq:appendix:rplusminus}
\end{align}

\begin{figure*}[ht!]
\begin{center}
\includegraphics[width=0.95\linewidth]{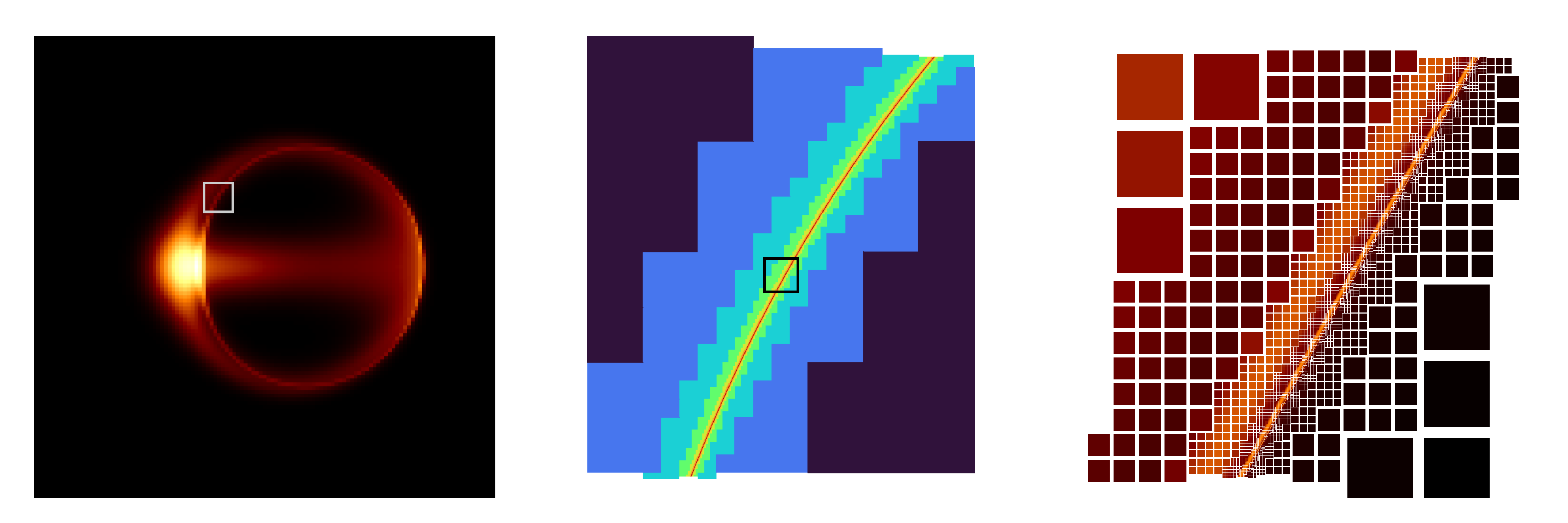} 
\end{center}
\caption{Left panel: Composite synthetic black hole image rendered on uniform grid composed by summing all subpixels within lowest-resolution image grid pixels. Center panel: Degree of refinement in region near the critical curve; full panel corresponds to white frame in left panel. Right panel: Pixel-centered intensities shown for pixel at all refinement levels, showing exponentially shrinking self-similar subring structure; full panel area corresponds to black frame in center panel.  }
\label{fig:appendix:numerics}
\end{figure*}

It is convenient to introduce a new variable for the latitudinal coordinate $\theta$. 
Substituting (\ref{eq:appendix:Phidefn}) and (\ref{eq:appendix:Qdefn}) and writing $u \equiv \cos \theta$, Equation~\ref{eq:thetadot} can be written as a fourth order polynomial in $u$: 
\begin{align}
    \Sigma^2 \dot{\theta}^2 &= Q - \left( \Phi^2 / \sin^2 \theta - \bhspin^2 \right) \cos^2 \theta \\
    \Sigma^2 \dot{u}^2 &= Q - \left( Q + \Phi^2 - \bhspin^2\right) u^2 - \bhspin^2 u^4 \\
    &= - \bhspin^2 \left( u^2 - u_+^2 \right) \left( u^2 - u_-^2 \right) . \label{eq:udotpolynomial}
\end{align}
The roots of the above expression
are given by (see also Equation~\ref{eq:upmsqdefn})
\begin{align}
    u_{\pm}^2 &\equiv \dfrac{\bhspin^2 - Q - \Phi^2 \pm \sqrt{\left(Q + \Phi^2 - \bhspin^2\right)^2 + 4 \bhspin^2 Q}}{2 \bhspin^2} .
\end{align}
In terms of $u$, the two other equations of motion are
\begin{align}
    \dot{t} &= \dfrac{ \left(\bhspin^2+r^2\right)^2 - \bhspin^2 \Delta \left(1-u^2
    \right)-2 \bhspin \Phi r}{\Delta \Sigma} \\
    \dot{\phi} &= \dfrac{2 \bhspin r+\Phi \left(r^2 + \bhspin^2 u^2  -2 r\right) \left(1-u ^2
    \right)^{-1}}{\Delta \Sigma} .
\end{align}

To compute the $\Delta \phi$ azimuthal precession identified in \S\ref{sec:reviewkerr}, we integrate $\ud \phi/ \ud u$ over four quarter cycles
\begin{align}
    \Delta \phi &= 4\int\limits_{0}^{\theta_+} \dfrac{\ud \phi}{\ud \theta} \, \ud \theta = 4 \int\limits_0^{u_+} \dfrac{\ud \phi}{\ud u} \, \ud u \nonumber \\
    &= \dfrac{4}{\sqrt{-u_-^2}} \left( \dfrac{2 r - \bhspin \Phi}{\Delta} K\left( \dfrac{u_+^2}{u_-^2}\right) + \dfrac{\Phi}{\bhspin} \Pi \left( u_+^2, \dfrac{u_+^2}{u_-^2} \right) \right) .
\end{align}
Here, we have written the answer in terms of complete elliptic integrals by expressing $\ud u$
in terms 
of the product written in Equation~\ref{eq:udotpolynomial}. The $\Delta t$ integral reported in Equation~\ref{eq:timedelaychi} is solved in the same way.

For $\bhspin \neq 0$, the ratio $\Delta \phi / 2\pi$ cannot be one, and thus a complete azimuthal cycle will not correspond to a complete latitude cycle. The sign of $\Delta \phi$ 
corresponds to the net displacement of the orbit (as either prograde or retrograde) and mirrors the sign of $\Phi$. 
The radius of the polar orbit, which has $\Phi = 0$, is given by \citep[e.g.,][]{Teo2003} 
\begin{align}
    r = 1 + 2 \sqrt{1-\dfrac{1}{3} \bhspin^2} \cos \left( \dfrac{1}{3} \arccos \dfrac{1 - \bhspin^2}{\left(1 - \frac{1}{3} \bhspin^2\right)^{3/2}} \right) \label{eq:appendix:Phi0}.
\end{align}

\section{Numerical image generation}

The ray traced results presented in this paper were produced using a custom version of the {\tt{}ipole} code \citep{Moscibrodzka2018}.
Most observer-to-emitter codes like {\tt{}ipole} solve the radiation transport equations along a single geodesic per resolution element (pixel). 
This approach is reasonable when the difference between neighboring geodesic trajectories is small (such as when the pixels are small or when the geodesics do not pass close to the photon sphere).
Since we study the neighborhood of the critical curve, we must resolve differences between geodesics as they begin to wind around the hole. Thus in our case, the pixel-centered method with a fixed grid fails since the widths of the subrings decrease exponentially. We deal with this issue by adaptively concentrating resolution elements near the critical curve.

The modified code first identifies the geodesics that have the longest path lengths relative to their neighbors and then constructs a connected set of pixels (the set containing the identified geodesics) to refine.
Each pixel is refined into a 3x3 set of pixels centered around the original geodesic. The process is repeated until a stopping criterion is met. 
Figure~\ref{fig:appendix:numerics} shows: the image produced by ray tracing on a grid with eight refinement levels, a schematic of refinement levels near the critical curve, and the multiple self-similar subrings produced by each of the imaged turnings.

\FloatBarrier

\bibliography{main}
\bibliographystyle{apj}

\end{document}